\documentclass[letter]{aastex63}
\usepackage[version=4]{mhchem}

\received{\today}
\revised{\today}
\accepted{\today}

\submitjournal{ApJ}

\shorttitle{Water UV-shielding and water emission}
\shortauthors{Bosman, Bergin, Calahan \& Duval}

\graphicspath{{./}{figures/}}

\begin{document}

\title{Water UV-shielding in the terrestrial planet-forming zone: Implications from water emission}

\correspondingauthor{Arthur D. Bosman}
\email{arbos@umich.edu}

\author[0000-0012-3245-1234]{Arthur D. Bosman}
\affiliation{University of Michigan, LSA astronomy \\
1085 S University \\
Ann Arbor, MI 48109, USA}

\author[0000-0003-4179-6394]{Edwin A. Bergin}
\affiliation{University of Michigan, LSA astronomy \\
1085 S University \\
Ann Arbor, MI 48109, USA}

\author[0000-0002-0150-0125]{Jenny Calahan}
\affiliation{University of Michigan, LSA astronomy \\
1085 S University \\
Ann Arbor, MI 48109, USA}

\author{Sara E. Duval}
\affiliation{University of Michigan, LSA astronomy \\
1085 S University \\
Ann Arbor, MI 48109, USA}

\begin{abstract}
Mid-infrared spectroscopy is one of the few ways to observe the composition of the terrestial planet forming zone, the inner few au, of proto-planetary disks. The species currently detected in the disk atmosphere, for example \ce{CO}, \ce{CO2}, \ce{H2O} and \ce{C2H2}, are theoretically enough to constrain the C/O ratio in the disk surface. However, thermo-chemical models have difficulties in reproducing the full array of detected species in the mid-infrared simultaneously. In an effort to get closer to the observed spectra, we have included water UV-shielding as well as more efficient chemical heating into thermo-chemical code Dust And Lines. We find that both are required to match the observed emission spectrum. Efficient chemical heating, in addition to traditional heating from UV photons, is necessary to elevate the temperature of the water emitting layer to match the observed excitation temperature of water. We find that water UV-shielding stops UV photons from reaching deep into the disk, cooling down the lower layers with higher column. 
These two effects create a hot emitting layer of water with a column of 1--10$\times 10^{18}$ cm$^{-2}$. This is only 1--10\% of the water column above the dust $\tau=1$ surface at mid-infrared wavelengths in the models and represents $< 1$\% of the total water column. 

\end{abstract}

\keywords{proto-planetary disks -- astrochemistry -- line formation}

\section{Introduction} \label{sec:intro}

The \textit{Spitzer Space Telescope} has revealed that the inner $\sim$1 au of most proto-planetary disks are rich in water and small organic molecules \citep[e.g.][]{Carr2008, Salyk2011, Pontoppidan2014PPVI}. It is within this same 1 au of the star a significant amount of the best studied exo-planets currently reside \citep[e.g.][]{Madhusudhan2019, Fulton2021}. Depending on if the planets have migrated to their current location or if they formed locally, there must be strong connection between the gas observed in the mid-infrared with the composition of these inner planets. 

To fully exploit this connection, and possibly unravel the formation origin of these inner planets, the elemental composition of the inner disk gas needs to be measured. After more than a decade of efforts on both the modeling and observational side, it is still unclear how to extract the elemental composition from the mid-infrared spectral information. The main carbon and oxygen carriers, CO, \ce{CO2} and \ce{H2O} can be observed in the infrared and so C/O ratios should readily attainable\citep{Najita2003, Carr2008, Pontoppidan2010, Salyk2011, Brown2013}. However, due to uncertainties associated with extracting accurate column densities from emission, it is difficult to derive accurate C/O ratios. Observation show that \ce{CO2} is not a dominant carrier of either Carbon or Oxygen \citep[][]{Pontoppidan2014, Bosman2017}. Detailed models of water get stuck on a degeneracy between gas-to-dust abundance and water abundance \citep{Meijerink2009, Blevins2016}. Whereas models of the \ce{CO} rovibrational lines are very sensitive to the assumed structures \citep{Bosman2019CO, Antonellini2020}. As such C/O ratios have only been inferred from infrared data in TW Hya, where \ce{H2} lines are available to infer the total column and inner disk structure can be resolved \citep{Bosman2019TWHya}.

An additional problem for extracting column densities from the observations is that the inferred column density from 2D thermochemical models and 1D slab models when reproducing the same spectra can differ by orders of magnitude. For water this disparity is best illustrated by comparing \citet{Meijerink2009} and \citet{Salyk2011}. The 2D models from \citet{Meijerink2009} predict \ce{H2O} that columns of $\gtrsim 10^{20}$ cm$^{-2}$ above the dust photo-sphere are necessary to reproduce the \textit{Spitzer}-IRS water spectra, while slab model analysis from \citet{Salyk2011} finds most disks have water columns between $10^{18}$ and $10^{19}$ cm$^{-2}$. More recent modeling efforts support the \citet{Meijerink2009} result of high \ce{H2O} columns above the dust photo-sphere \citep{Blevins2016, Woitke2019}. Similar column discrepancies between slab and 2D models are also seen for \ce{HCN} \citep{Bruderer2015} and \ce{CO2} \citep{Bosman2017}. This uncertainty in the conversion of observed molecular column to  makes it difficult to properly anchor any attempt at extracting C/O ratios directly from the column densities of \ce{CO} and \ce{H2O}. 

Modeling studies have shown that the composition of gas within the inner disk is dependent on the elemental composition specifically the C/O ratio \citep{Woitke2019, Anderson2021}. This could allow for a C/O ratio measurement that is independent of the absolute column of \ce{H2O}. However, current models cannot simultaneously match \ce{H2O}, \ce{CO2} and \ce{C2H2}, the three species most sensitive to the C/O ratio. This indicates that current models are missing part of the thermo-chemical puzzle. In particular the effects of water UV-shielding have been shown to be present \citep{Bethell2009}, but are not widely included in modeling efforts, furthermore excess heating of the gas is often invoked \citep[e.g.][]{Glassgold2009, Meijerink2009, Glassgold2015, Anderson2021}, but seldom consistently included a full thermo-chemical model. 
In particular the models of \citet{Adamkovics2014} and \citet{Adamkovics2016} do include these effects. However, they assumed a gas-to-dust ratio of 100 in disk surface layers which would not allow for the high \ce{H2O} columns inferred from the infrared spectra \citep{Meijerink2009, Blevins2016, Woitke2019}. As a result it is unclear if these models would reproduce the observed emission.

This paper is the first in a series of four (Bosman et al. in prep, Calahan et al in subm, Duval et al in prep) that explores these issues and presents a way forward for the interpretation of infrared spectral data. The current paper will focus on water itself, using a state-of-the-art thermo-chemical model to investigate the effects of additional heating as well as water UV-shielding on the predicted water spectra. Future papers will focus on the \ce{CO2}-\ce{H2O} ratios (Bosman et al. in prep), Lyman-$\alpha$ and \ce{H2{}^{18}O} (Calahan et al. in subm) and organics (Duval et al. in prep).

\section{Methods}

\begin{figure*}
    \centering
    \includegraphics[width = \hsize]{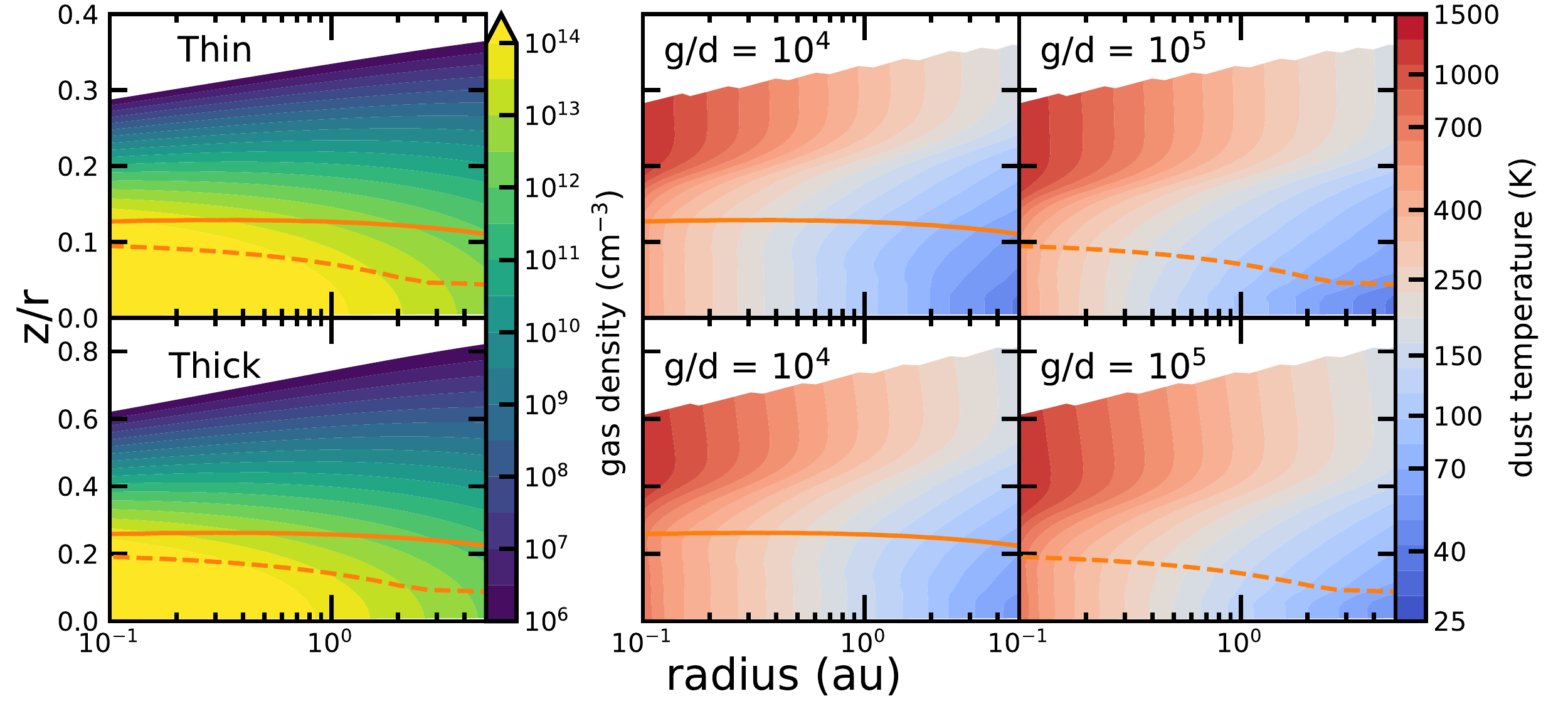}
    \caption{Gas density (left) and dust temperature (right) for the different gas and dust structures. Top row shows the thin model ($h_c = 0.08$) and the bottom row the thick model ($h_c = 0.16$). The gas-to-dust ratio in the surface layers after settling the large dust is denoted in the top left of the dust temperature panels. Orange lines show the continuum $\tau=1$ line at 15$\mu$m for the g/d = $10^4$ (solid) and $10^5$ (dashed) dust distributions. }
    \label{fig:dust_dens_temp}
\end{figure*}

Our models are based on the Dust and Lines (DALI) code \citep{Bruderer2012,Bruderer2013}. With a number of extensions to better model the hot inner regions of the disk. The chemical network has been expanded in two ways: by increasing the formation of \ce{H2} and including water UV-shielding. 

As noted by \citet{Glassgold2009}, the presence of water vapor in the warm to hot ($\sim$few hundred K to 1000~K) disk atmosphere requires the a priori presence of \ce{H2}. Thus in our model we have updated the chemistry, especially the formation of \ce{H2}, as described in App.~\ref{app:reactions}. Specifically 3-body formation reactions have been added and \ce{H2} formation on grains at temperatures between 300 and 900 K has been increased in line with experiments \citep[][see also \citet{Thi2020}]{Cazaux2002, Cazaux2004,Wakelam2017}.

The inferred \ce{H2O} columns ($>10^{18}$ cm$^{-2}$) with Spitzer-IRS imply that enough \ce{H2O} exists in the inner disk to contribute significantly to the UV opacity of the inner disk gas as noted by \citet[][]{Bethell2009}. Therefore we include shielding of UV photons by \ce{H2O} \citep[Using cross sections from][]{Chan1993, Fillion2003, Fillion2004, Mota2005, Heays2017}. We include effect of \ce{H2O} on the UV flux propagation as a vertical extinction term during the chemistry and thermal balance. As such we do not just include a self-shielding term, as is common for line dominated species such as \ce{CO}, \ce{N2} and \ce{H2}, but a full UV-shielding term, which impacts the dissociation rate of all species as well as the amount of energy injected into the gas by UV photons. The \ce{H2O} cross section is averaged within the wavelength bins used during the calculation and the UV flux in the cell is updated based on the \ce{H2O} column above the cell.

To be able to self-consistently test the effect of additional heating on the chemistry and line emission, we examine two gas heating formalisms. The standard DALI heating \citep{Bruderer2012, Bruderer2013} and extra chemical heating due to photo-dissociation and molecule formation, following the high density results from \citep{Glassgold2015}. 

The physical model is a smooth version of the AS 209 model from \citet{Zhang2021MAPS}. The disk parameters are given in Table~\ref{tab:All_mod_param}. The stellar input spectrum is also taken from \citet{Zhang2021MAPS} and is the combination between a stellar atmosphere model \citep[Nextgen, ][]{Hauschildt1999} and excess UV \citep{Herczeg2004, Dionatos2019}, which is dominated by Ly-$\alpha$ emission, as consistent with observations \citep[][]{Schindhelm2012}. The final stellar spectrum is spectral type K5 (4300 K) with a total luminosity of 1.4 $L_\odot$ and 0.01 $L_\odot$ in UV at wavelengths smaller than 200 nm.  AS 209 has been chosen as a base model, as its accretion rate of $10^{-7.5}$ $M_\odot$ yr$^{-1}$ creates a more typical steller irradiation environment than an AS 205N inspired model \citep[high UV e.g.,][]{Bruderer2015, Bosman2017} or TW Hya inspired model \citep[low UV;][]{Woitke2018, Anderson2021}. The disk mass in the model, 0.0045 $M_\odot$, is also more typical than the $\gtrsim 0.01$ $M_\odot$ disk masses of the aforementioned studies \citep[e.g.][]{vanTerwisga2022}. 

As the physical structure is very \textit{thin}, with a scale-height of 0.05 at 1 AU ($h_c = 0.08$), we have also included a \textit{thick} model with larger vertical distribution, yielding a scale height of 0.1 at 1 AU ($h_c = 0.16$). 
Most of the dust is assumed to be in large dust settled towards the mid-plane, yielding gas-to-dust ratios in the surface layers of $10^4$ (99\% large, settled dust) and $10^5$ (99.9\% large, settled dust). Whereas many previous studies have assumed large grains (up to 1 mm) \citep[e.g.][]{Bruderer2015, Bosman2017, Woitke2018} we only use small grains (up to 1 $\mu$m) in the inner disk surface. Dust opacities are calculated using in DSHARP dust opacity tool \citep{Birnstiel2018} and are the same as the ones used in \citet{Zhang2021MAPS}. 

The water emission lines are calculated using the molecular data file from the Leiden Atomic and Molecular Database. Levels with energies up to 7200 K are included \citep{Tennyson2001}. Line transitions are taken from the BT2 list \citep{Barber2006} and the collisional rate coefficients are from \citep{Faure2008}. The water spectra are calculated from the non-LTE level populations using the ``fast line ray tracer'' as described in \citep[][App B.]{Bosman2017}. 

To see how our models compare with observations, we create a representative slab model using the parameters in \citet{Salyk2011}. We use a water excitation temperature of 500 K and a water column of $3\times 10^{18}$ cm$^{-2}$. The spectra are calculated using the slab model from \citep{Banzatti2012}. Finally, the spectrum is scaled to have the same emitting area as the model with both water UV-shielding and extra chemical heating.

\begin{table}
\centering
\caption{\label{tab:All_mod_param} Model parameters}
\begin{tabular}{l c c}
\hline
\hline
Parameter & Symbol&  Value \\
\hline
Stellar Luminosity & &1 $L_\odot$ \\
Stellar Spectrum & & AS 209 $^{a}$\\
Stellar Mass & & $ 1.0 M_\odot$ \\
Sublimation radius & $R_\mathrm{subl}$ & 0.08 AU \\
Critical radius &$R_c$ & 46 AU \\
Disk outer radius & $R_\mathrm{out}$ & 100.0 AU \\
Gas surf. dens. at $R_c$ & $\Sigma_c$ & 21.32 g cm$^{-2}$ \\
Surf. dens. powerlaw slope & $\gamma$ & 0.9 \\
Disk opening angle & $h_c $ & [0.08, 0.16] \\
Disk flaring angle & $\psi$ & 0.11 \\
Large dust fraction& & [0.99, 0.999] \\
Large dust settling& $h_d/h_g$ & 0.2 \\
\hline
\end{tabular}
\tablecomments{$^{a}$  \citep{Zhang2021MAPS} }
\end{table}

\begin{figure*}

    \begin{minipage}{0.499\hsize}
    \includegraphics[width = \hsize]{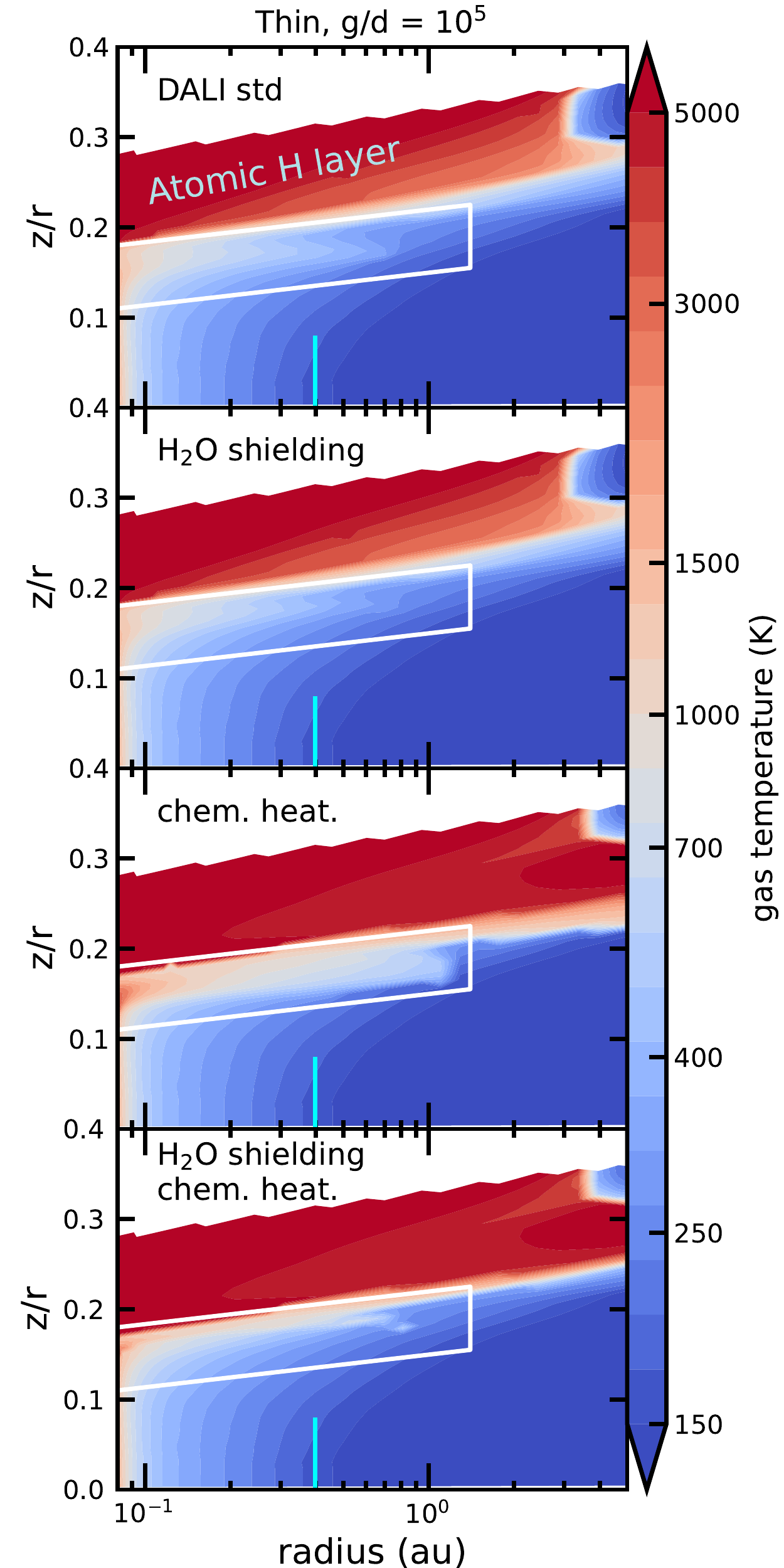}
    \end{minipage}%
    \begin{minipage}{0.499\hsize}
    \includegraphics[width = \hsize]{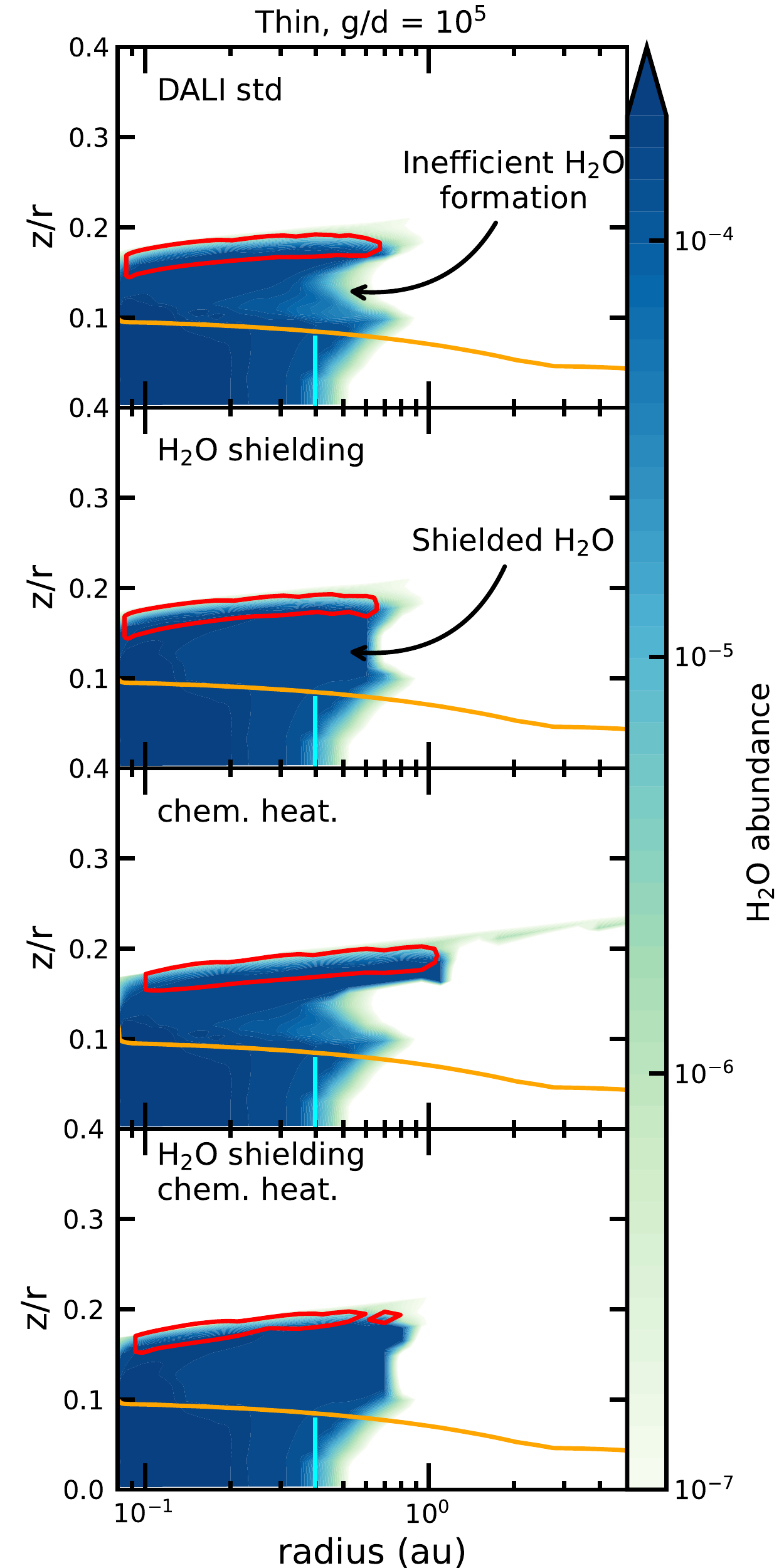}
    \end{minipage}
    \caption{Gas Temperature (left) and water abundance (right) for the thin model with a surface layer gas-to-dust ratio of $10^5$ for variations in the chemical and heating-cooling models. The vertical blue lines show the mid-plane \ce{H2O} snowline location in the model. Above a $z/r$ of $\sim$ 0.2 hydrogen is dominantly in atomic form and the gas is hot ($>$1000 K). Below this models show a warm layer where the gas and dust temperatures are strongly decoupled, but the gas is molecular (region enclosed in the white box). This layer is critical for the inner disk emission. The gas in this layer is heated by UV photons, as such, water UV-shielding cools down this layer, while extra chemical heating leads to higher temperatures. 
    In the right panels, the orange line shows the continuum $\tau=1$ line at 17 $\mu$m and the red contours show the origin of 90\% of the $11_{3,9}$--$10_{0,10}$  $17.2 \mu$m water line flux. Including the effect of water UV-shielding strongly increases the abundance of \ce{H2O} between a z/r of 0.1 and 0.15 at radii $>$0.3 au. }
    \label{fig:Tgas_chem}

\end{figure*}

\section{Results}

\subsection{Temperature structure}

Figure~\ref{fig:dust_dens_temp} shows the gas density and dust temperature for the four different gas and dust structures. The thin models are generally cooler than the thick models near the mid-plane. This difference disappears in the upper layers where the unattenuated stellar radiation field dictates the dust temperature. 
The models with a larger surface layer gas-to-dust ratio are also slightly cooler in the mid-plane as heat escapes more easily in the vertical direction. The larger surface layer gas-to-dust ratio also pushes down the vertical temperature transition, between the heated surface layer and the mid-plane. This is a result of lower continuum opacities at all wavelengths. 

Figure~\ref{fig:Tgas_chem} presents the gas temperature structure for the thin model with a gas-to-dust ratio of 10$^5$ in the disk surface for four different thermo-chemical iterations: standard model (\textbf{DALI std}), standard model with water UV-shielding (\textbf{\ce{H2O} shielding)}, standard model with chemical heating (\textbf{chem. heat.}), and standard model with both water UV-shielding and chemical heating (\textbf{\ce{H2O} shielding, chem. heat.}). Appendix \ref{app:tgasthick} discusses effect of the different structures. The general behavior seen in Fig.~\ref{fig:Tgas_chem} are also seen in the other models. 

In general, around the location of the dust temperature transition from the heated surface, to close to the mid-plane temperature is where gas and dust become strongly coupled. Above this vertical point (z/r $\gtrsim$ 0.15) is where gas and dust are thermally decoupled and where changes in the thermo-chemistry lead to changes in the gas temperature. 

A small effect of water UV-shielding can be seen in the difference between the models with and without water UV-shielding between a z/r of 0.15 and 0.2 (white box in Fig.~\ref{fig:Tgas_chem}). The water UV-shielding models have lower gas temperatures in this region as less UV photons reach this layer. This provides less heating of the gas as a result of photo-dissociation and molecular formation. 
When extra chemical heating following photo-dissociation is included, the molecular layer at $z/r \sim 0.15$--$0.2$ is significantly heated, nearly doubling in gas-temperature. This region is significantly larger in the case where water UV-shielding is not included. This region is fully molecular and will therefore  have a significant impact on the resulting emission line spectra. 

\subsection{Water abundance}

Figure~\ref{fig:Tgas_chem} also provides the water vapor abundance structure for the thin model with a gas-to-dust ratio of 10$^5$ in the disk surface for the same four different thermo-chemical iterations. Different choices in the thermo-chemical model have strong implications for the water abundance structure. The base DALI model shows a gap in the \ce{H2O} abundance between a z/r of 0.1 and 0.15. In this relatively cold gas, $<500$K, formation of \ce{OH} from \ce{O + H2} is very slow. This allows the attenuated UV field in this region to keep a large fraction of the oxygen in atomic form. 

This low water abundance region is no longer present when the UV-shielding of water is included. The self-shielding effect filters out all the \ce{H2O} dissociating photons in surface layers allowing \ce{H2O} to survive in the deeper, colder layers. There now exists a very sharp drop in the \ce{H2O} abundance around 0.6 au when the surface layer is not warm enough to efficiently form water and no shielding layer exists. Outside this point most of the oxygen in the model is in atomic form. When the dust becomes cold enough for water to freeze-out, water ice dominates the oxygen budget. Additional chemical heating has little effect on the water vapor abundance.  However, the higher gas temperature produced in the surface extend the water-rich layer outward.

\begin{figure*}
    \centering
    \begin{minipage}{0.49\hsize}
    \includegraphics[width = \hsize]{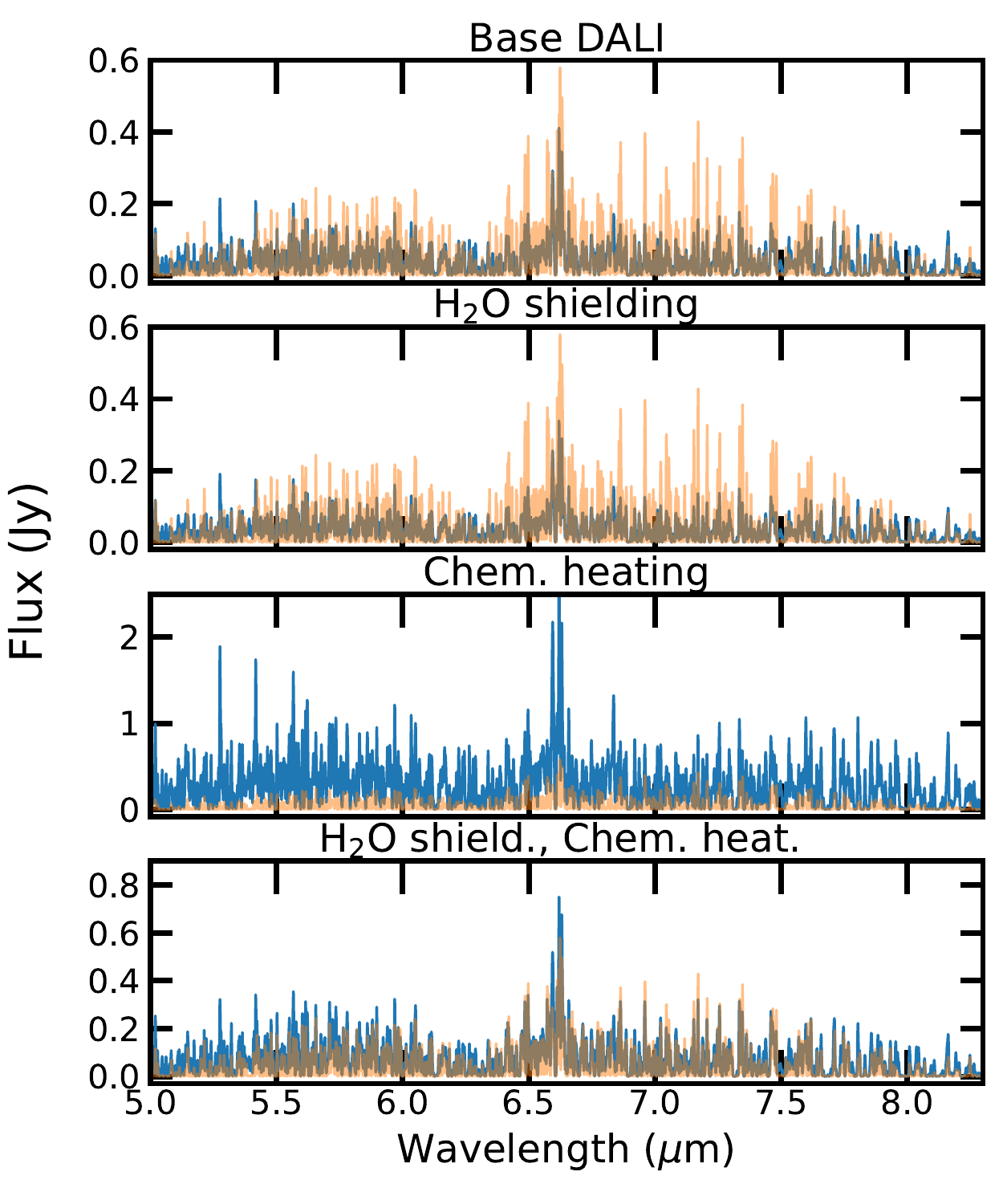}
    \end{minipage}%
    \begin{minipage}{0.49\hsize}
    \includegraphics[width = \hsize]{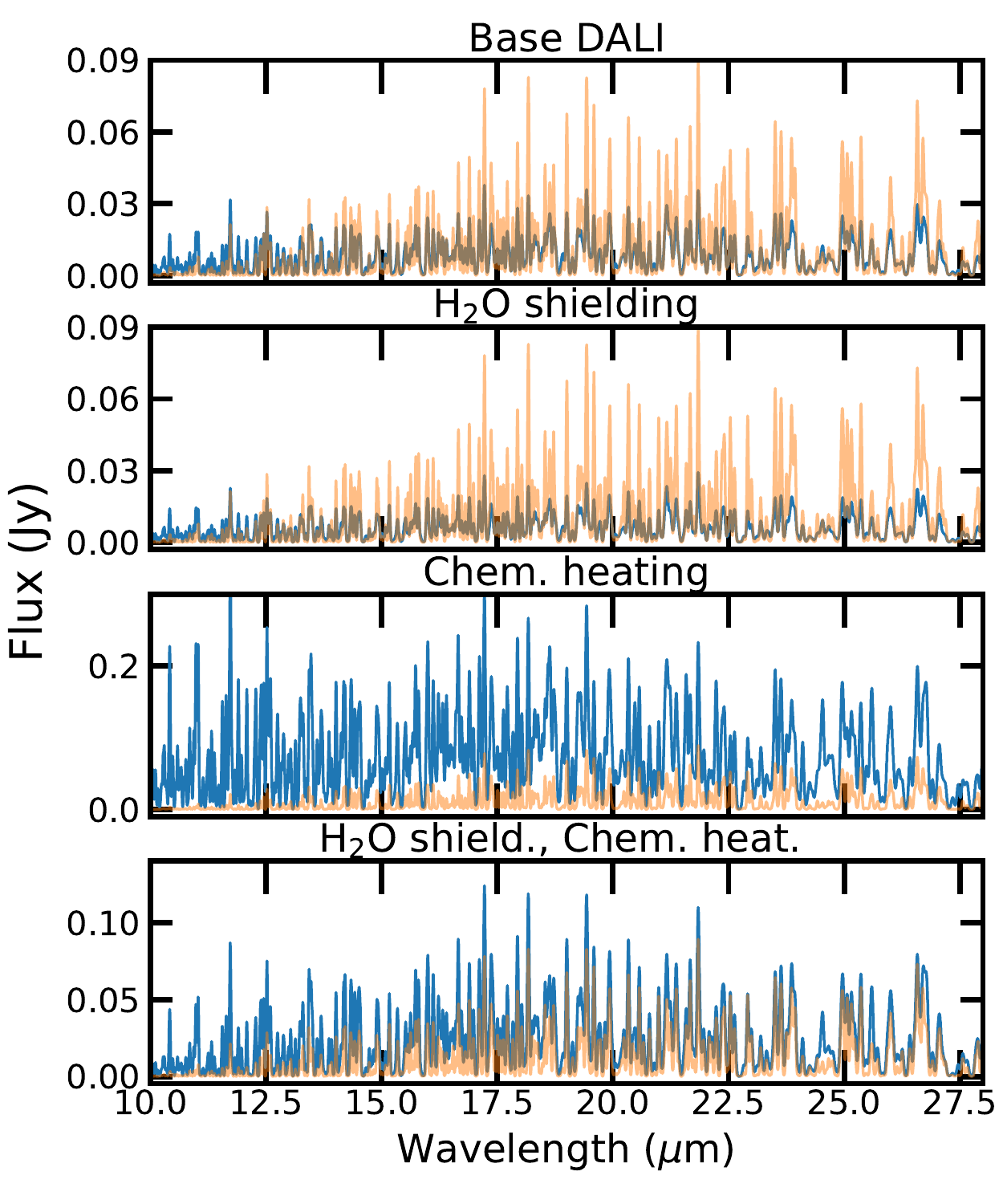}
    \end{minipage}

    \caption{Water spectrum over most of the JWST-MIRI observable range for the thin model with a gas-to-dust ratio of $10^5$ in the surface layers (blue). The spectrum between 5 and 8 micron (left) is convolved to a resolving power of R = 3000, comparable to MIRI, the 10-28 micron spectrum on the right is convolved to an R = 600, to ease comparison with \textit{Spitzer} spectra. The orange spectrum shows a water spectrum as computed from a typical slab model \citep{Salyk2011}, with an excitation temperature of 500 K and a \ce{H2O} column of $3\times 10^{18}$ cm$^{-2}$. In the \textit{Spitzer} range the model standard DALI and water UV-shielding models under predict the \ce{H2O} flux, except for the 10-12 micron regions. The extra heating models, however, overestimate the \ce{H2O} flux, but the model with both extra heating and water UV-shielding fits the slab model better than the model with just water UV-shielding. }
    \label{fig:Spectra_thinE5}
\end{figure*}

\subsection{Water spectra}

Figure~\ref{fig:Spectra_thinE5} show the water spectra over the MIRI range compared to a typical slab model \citep{Salyk2011}. The slab model is the best fit to the observed Spitzer/IRS water spectrum and thus is an effective reproduction of the observed water emission.  In most of the 10-28 $\mu$m range our models without additional heating underestimate the line flux.  This is the case even though the emitting area is the same for the slab and 2D model; this indicates that higher temperatures are necessary. Including chemical heating without water UV-shielding creates a spectrum that is very bright due to the gas heating over a large column. 

The inclusions of water UV-shielding confines the heating to a thin column and results is a good match to the slab spectrum.

We note that the brightness ratio between shorter wavelength flux and longer wavelength flux for the pure rotational lines is always higher in the DALI models compared to the slab models.
Interestingly when comparing the vibrational 6.5 $\mu$m feature and the 10-28 $\mu$m rotational lines, the opposite relation is seen. From the comparison between the slab model and DALI model spectrum at around 11 $\mu$m, it would be expected that the DALI 6.5 $\mu$m feature would be brighter in the all DALI models. This is only seen in the chemical heating model without water UV-shielding, which, as already discussed, is way too bright in the longer wavelengths as well. This indicates that water emission is not fully in LTE over the MIRI band, which we discuss in Sec.~\ref{sec:disc_excite}. Spectra for the different dust structures are discussed in App.~\ref{app:dust}.

\begin{figure}
    \centering
    \includegraphics[width = \hsize]{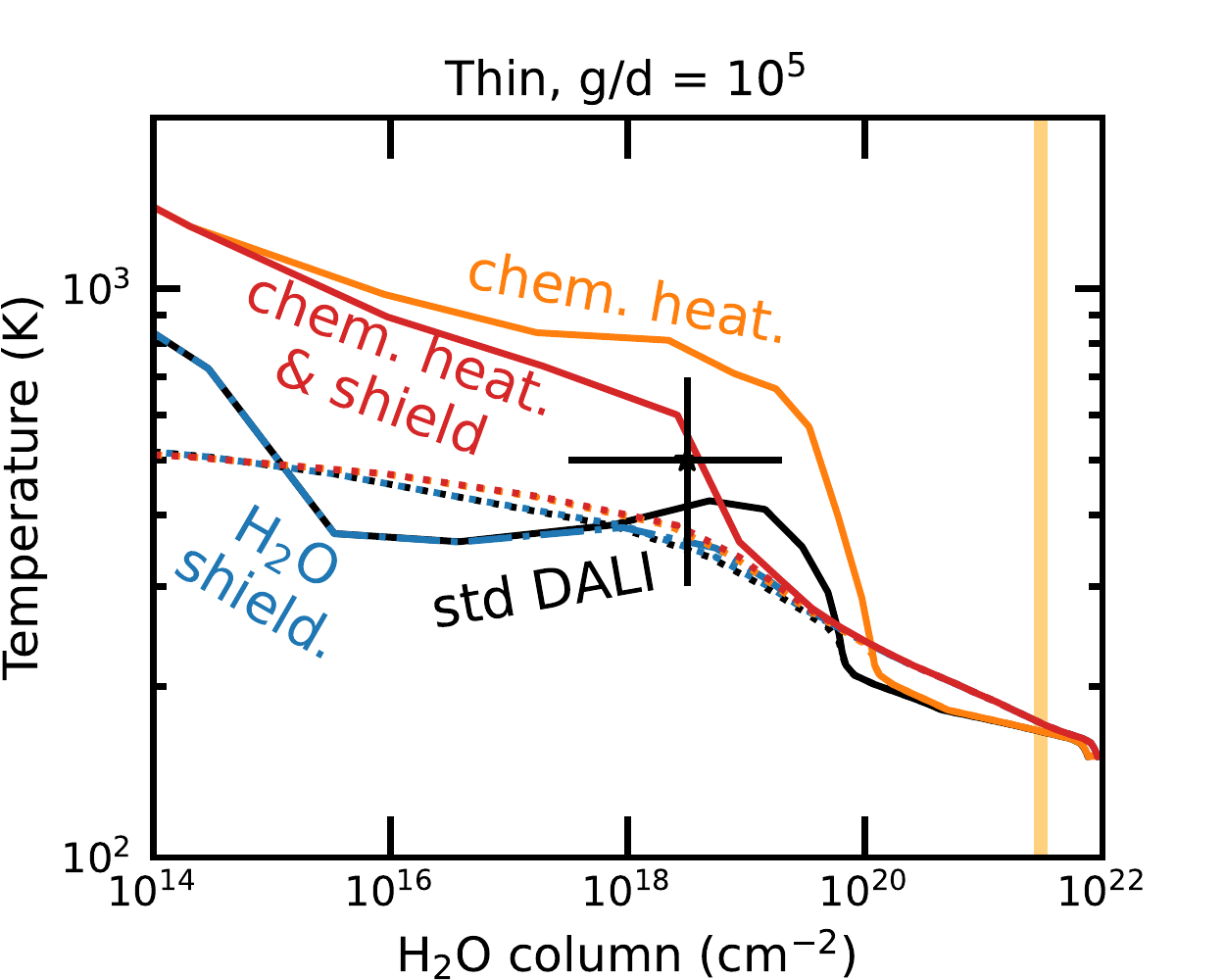}
    \caption{Gas temperature as function of the cumulative vertical \ce{H2O} column at the location of the mid-plane \ce{H2O} iceline (0.4 au). The \ce{H2O} column is a proxy for the location in the disk, higher column being deeper into the disk. Due to the different abundance profiles in the models these \ce{H2O} columns map to different total gas columns for the different models. Solid and dash-dotted lines show the gas-temperature profiles and the dotted lines show the dust temperature. 
    The black star with error bar gives the range of columns and excitation temperatures extracted from the \textit{Spitzer}-IRS spectra \citep{Salyk2011}. The vertical orange line shows the approximate location where the dust becomes optically thick at 17$\mu$m. Up to a column of $10^{17}$ cm$^{-2}$ the standard DALI and the water UV-shielding models are indistinguishable. At higher water columns, the shielding of \ce{H2O} kicks in and heating by UV photons is suppressed. }
    \label{fig:Temp_col_relation}
\end{figure}

\section{Discussion}

\subsection{Inner disk temperature structure}

The models clearly show that both the assumptions on chemical heating as well as the UV attenuation due to \ce{H2O} has a strong impact on the predicted gas temperature and water abundance structure and resulting \ce{H2O} spectra. Both should thus be considered in thermo-chemical models. 

The water main water emitting layer (\ce{H2O} columns of $\sim 10^{18}$ cm$^{-2}$ or more) are at high densities, $>10^{10}$ cm$^{3}$, as such in these layers, the very dense approximation of \citet{Glassgold2015} should hold. Above this layer, the heating might be overestimated, as more energy can escape radiatively. This gas is not contributing to the emission so this is not critical for our conclusions. 

The inclusion of water UV-shielding creates a region of the molecular layer that is shielded from UV photons by water and is also cooler. This layer starts below a water column of a few times $10^{18}$ cm$^{-2}$, corresponding on a \ce{H2} column of 5--30 $\times 10^{22}$ cm$^{-2}$. This jump in temperature should be observable, as cold lines, with a large column, and hotter lines, coming from a smaller column in various inner disk tracers. 

The rotational water spectra also contains information of the radial temperature gradient. The model water spectra are clearly not well captured by a single temperature slab model. The relative brightness of the 10-14 $\mu$m lines compared to the slab model prediction is due to the higher temperatures at smaller radii. These boost the higher excitation lines more than the fall-off due to emitting area. All our models, regardless of the inclusion of excess heating show a power-law temperature-radius relation with a coefficient $q \sim -0.7$. Slab model fitting with this temperature profile will be necessary to properly describe the model, and most likely, the observed purely rotational lines of \ce{H2O}. 

\subsection{Water Excitation}
\label{sec:disc_excite}
The vibrational band at 6.5 $\mu$m will require a similar radial rotational temperature profile as the pure rotational line. However, the comparison between the DALI model and slab spectra show that this will not be enough, with the 6.5 $\mu$m being relatively weak. This mismatch in line brightness is due to the excitation of water. The highly excited rotational lines that dominate the spectrum longward of 10 $\mu$m are close to LTE in the emitting region (densities of $10^{11}$--$10^{12}$ cm$^{-3}$), however at these densities, the vibrational band at 6.5 $\mu$m is not in LTE and is subthermally excited. Thermalisation in the model happens at densities of $\sim$ $10^{13}$ cm$^{-3}$. 

{\em The excitation of water emission as measured with JWST-MIRI can thus be used to estimate the density of the emitting region.} If the 6.5 $\mu$m band is weaker than expected from an LTE model fit to the 10-15 $\mu$m data, this is likely due to sub-thermal excitation of \ce{H2O} and thus (relatively) low density of the gas. This should be robust against uncertainties of the emitting area as the upper-level energies of the lines around the center of the vibrational band ($\sim$ 2300 K) are similar to the upper level energies of the lines around 12 $\mu$m ($\sim$ 2400 K).

\subsection{The inner disk water reservoir}

Infrared line observations only probe the surface layers of the disk. To be extrapolate these observations to more general statements of the inner disk, both in the terms of chemical complexity as well as in terms of midplane physical conditions it is critical to understand what part of the disk we are actually probing with in infrared spectra. 

Figure~\ref{fig:Tgas_chem} also illustrates the primary emission zone
of water vapor at mid-IR wavelengths.
The chemical models naturally and strongly predicts a contained emission region that extends up to twice the midplane water snowline radius. This agrees well with the analysis by \citet{Meijerink2009} and \citet{Blevins2016} that the water emission in T-Tauri disks comes from a strongly radially-contained region. With the variation in the observation data and in our model predictions, it is not possible to say if this containment is strong enough, or if containment to within the radius of the water mid-plane snowline is necessary \citep[as proposed by][]{Meijerink2009, Bosman2021water}.

Vertically most of the water emission is found arising from a thin hot layer where the gas temperature structure is set by water UV-shielding.  This zone is well above the dust optical surface.
The layer probed by water infrared emission, and probably by many other inner disk tracers is thus not set by the dust optical depth at the observed wavelength. Water lines with low Einstein A coefficient lines or \ce{H2O} isotopologues will be able to probe the deeper water abundance structure (see, e.g. Calahan et al. in prep). It is these deeper probing lines that that will be necessary to constrain bulk disk surface properties, as the C/O ratio and the gas-to-dust ratio.

However, our models do provide some information on the water column probed by observations, which is distinct from the total water column in the inner disk.
Figure~\ref{fig:Temp_col_relation} shows the vertical gas and dust temperature structure as function of the \ce{H2O} column at the radius of the \ce{H2O} midplane compared to the column density estimated via slab models of the water vapor emission in T Tauri systems observed by Spitzer \citep[][]{Salyk2011}.  Overall our models find a much larger water column is present in the system at these radii (i.e. $\sim$10$^{22}$~cm$^{-2}$ compared to few $\times$ 10$^{18}$~cm$^{-2}$).   However, what matters for the emission lines for transitions associated with mid-IR wavelengths is the column that is present at high temperature and the optical depth of these lines. Here we have two points to make.  First, many of the strong waterlines are optically thick at a column of $10^{18}$ cm$^{-2}$.  Second, the gas temperature is rapidly falling below 400~K near a water column of a few $\times$ 10$^{18}$~cm$^{-2}$.  This effectively becomes the column traced by the mid-infrared lines.  This is despite the fact that the total column exceeds this value \citep[as first noted by][]{Meijerink2009}. 
This implies that the \textit{Spitzer} spectra are only probing a small fraction of the observable inner disk water reservoir.
In Calahan in prep. we illustrate that certain transitions of \ce{H2{}^{18}O} can gain access to this reservoir.  Regardless, detailed models will be needed to match observational data and infer the underlying hidden \ce{H2O} content.

\subsection{M-dwarf and Herbig Ae disks }
The model discussed so far are representative of disk around a young, solar mass star. However, many of the objects to be observed will have a far different mass range and spectral energy distributions. To capture some of these effects we looked at two additional model runs, one around an M-dwarf \citep[M3, 3300 K, 0.1 $L_\odot$, $L_{<\mathrm{200 nm}} = 10^{-3}L_\odot$]{Hauschildt1999, Herczeg2004} and Herbig Ae \citep[A1, 9300 K, 14 $L_\odot$, $L_{<\mathrm{200 nm}} = 0.8 L_\odot$, HD 163296 in][]{Zhang2021MAPS}. The X-ray luminosity is taken to be $10^{29}$ erg s$^{-1}$ for both models. The disk structures are identical to those from Table~\ref{tab:All_mod_param}, except that the sublimation radius is scaled to 0.022 and 0.3 au for the M-dwarf and Herbig Ae disk respectively. 

The M-dwarf disk water spectra in general are well described by the T-Tauri water spectra linearly scaled down with the stellar luminosity, especially for the models with a gas-to-dust ratio of $10^4$ in the disk surface layer. For a gas-to-dust ratio of $10^5$ the models with extra chemical heating about a factor 2 brighter than would be expected from a linearly scaled down T-Tauri model.

The Herbig Ae disk water spectra are across the board brighter than a T-Tauri model linearly scaled up with luminosity; likely due to the higher UV-to-total luminosity ratio of the Herbig Ae stellar spectrum. This causes a higher contrast between midplane dust temperature and surface layer gas temperature, leading to stronger water emission relative to the total stellar luminosity. This effect is super charged in the models with a gas-to-dust ratio of $10^5$ that include chemical heating. While the total luminosity increases by a factor of 10, the water line fluxes between 10 and 28 $\mu$m increase by more than a factor of 100. 

For the Herbig Ae disks, these stronger line fluxes (and stronger line-to-continuum ratios) are counter to the observations, were water is not detected towards to majority of sources \citep{Pontoppidan2010,Antonellini2016}. This implies that there is a fundamental difference in the disk structures between T-Tauri and Herbig Ae disks. This might be related lower gas-to-dust ratios in the surface layers of Herbig Ae disk compared to T-Tauris, leading to low contrast between water lines and the continuum \citep{Antonellini2016} or to the inferred puffed-up inner rims which could radially constrain the infrared lines emission, which is also seen in the CO ro-vibrational lines \citep{Dullemond2010,Bosman2019CO}.

\section{Conclusions}

We have studied the impact water UV-shielding and chemical heating on the emission of water in the mid-infrared. We find that the inclusion of water UV-shielding and extra chemical heating significantly impacts the temperature in the water line emitting layer. Both are require to match the observed spectra. The extra chemical heating raises the temperature in the top of the water emitting layer to the observed values. Water UV-shielding cools down the disk below this region by exclusion of the UV photons. This limits the \ce{H2O} column that can emit to the observed column values.  

This results in the models having a large region above the dust mid-infrared photo-sphere that contains colder water that does not significantly contribute to the observed spectrum. This leads to a 1--2 order of magnitude mismatch between the theoretically observable water column and the water column implied by the spectrum. This reservoir might be visible with select \ce{H2^{18}O} lines. Radially the \ce{H2O} emitting region is naturally confined by the chemistry. 

Model spectra are not fully captured by a single LTE slab model. High upper level energy lines generally originate from a smaller but hotter emitting region, while the vibrational emission around 6.5 $\mu$m comes from vibrationally sub-thermally excited \ce{H2O}, while the rotational transitions are generally in LTE.   This suggests that future observations from JWST could extract both density and temperature within emitting layers.

\acknowledgments 

The authors thank the referee for a constructive report that improved the quality of the paper. We also thank Beno\^it Tabone and Stephanie Cazaux for useful discussion on \ce{H2} formation.
ADB and EAB acknowledge support from NSF Grant\#1907653 and NASA grant XRP 80NSSC20K0259. 
\software{Astropy \citep{astropy2013,astropy2018}, SciPy \citep{Virtanen2020},  NumPy \citep{van2011numpy}, Matplotlib \citep{Hunter2007}, DALI \citep{Bruderer2012, Bruderer2013}}

\bibliography{Lit_list}{}
\bibliographystyle{aasjournal}

\appendix

\section{Reactions added to the network}
\label{app:reactions}
In our model \ce{H2} formation in warm gas has to be increased to create water abundances high up in the disk \citep{Glassgold2009}. This is facilitated by increasing the chemisorption binding energy from 10000~K to 30000~K within the formalism of \citep{Cazaux2002, Cazaux2004}. With these assumptions, the dust temperature allows \ce{H2} to form on the grains from $\sim$300 to $\sim$900~K in line with experiments of \ce{H2} formation in the lab on various surfaces \citep[][Cazaux, private communication]{Wakelam2017}; this is in line with assumptions made by \citep{Adamkovics2014}. As densities in the inner disk can reach $10^{12}$ cm$^{-3}$ in the UV penetrated layers, we also included a number of 3 body reactions. Notably, this included the 
\begin{equation}
 \ce{H + H + H -> H2 + H} 
\end{equation}
reaction which increase the \ce{H2} formation rate at higher density. A full list of the reactions and their coefficients can be seen in Table~\ref{tab:react}. Reactions are taken from the network used in \citet{Walsh2015}. 

\begin{table}[]
    \centering
    \caption{Reactions with reaction coefficients ($k = a \times (T/300)^b \exp(-c/T)$)}
    \begin{tabular}{l c c c | l c c c}
    \hline
    \hline
     Reaction & a & b  & c (K) &Reaction & a & b  & c (K)\\
    \hline
\ce{H     + H    + H   ->H2    + H         }      &      1.422(-32)  & -0.2  &  0         \\  
\ce{H     + H    + H   ->H2    + H         }      &      1.150(-32)  & -0.5  &  0         \\  
\ce{H     + H    + H2  ->H2    + H2        }      &      9.100(-33)  & -0.6  &  0         \\
\ce{H2    + e-         ->H     + H    + e- }      &      3.220(-09)  & -0.3  &  1.020(5)  \\
\ce{C     + H2   + H   ->CH2   + H         }      &      6.900(-32)  &  0    &  0        & \ce{C     + H2   + H2  ->CH2   + H2        }      &      6.900(-32)  &  0    &  0         \\
\ce{CH    + H2   + H   ->CH3   + H         }      &      5.100(-30)  & -1.6  &  0        & \ce{CH    + H2   + H2  ->CH3   + H2        }      &      5.100(-30)  & -1.6  &  0         \\
\ce{CH3   + H    + H   ->CH4   + H         }      &      6.300(-29)  & -1.8  &  0        & \ce{CH3   + H    + H2  ->CH4   + H2        }      &      6.300(-29)  & -1.8  &  0         \\
\ce{O     + H    + H   ->OH    + H         }      &      4.330(-32)  & -1.0  &  0        & \ce{O     + H    + H2  ->OH    + H2        }      &      4.330(-32)  & -1.0  &  0         \\
\ce{OH    + H    + H   ->H2O   + H         }      &      2.600(-31)  & -2.0  &  0        & \ce{OH    + H    + H2  ->H2O   + H2        }      &      2.600(-31)  & -2.0  &  0         \\
\ce{NH2   + H    + H   ->NH3   + H         }      &      3.010(-30)  &  0    &  0        & \ce{NH2   + H    + H2  ->NH3   + H2        }      &      3.010(-30)  &  0    &  0         \\
\ce{O     + O    + H   ->O2    + H         }      &      5.210(-35)  &  0    & -9.000(2) & \ce{O     + O    + H2  ->O2    + H2        }      &      5.210(-35)  &  0    & -9.000(2)  \\
\ce{N     + N    + H   ->N2    + H         }      &      1.380(-33)  &  0    & -5.030(2) & \ce{N     + N    + H2  ->N2    + H2        }      &      1.380(-33)  &  0    & -5.030(2)  \\
\ce{CO    + H    + H   ->HCO   + H         }      &      6.300(-35)  &  0.2  &  0        & \ce{CO    + H    + H2  ->HCO   + H2        }      &      6.300(-35)  &  0.2  &  0         \\
\ce{CN    + H    + H   ->HCN   + H         }      &      8.500(-30)  & -2.2  &  5.670(2) & \ce{CN    + H    + H2  ->HCN   + H2        }      &      8.500(-30)  & -2.2  &  5.670(2)  \\
\ce{CO    + O    + H   ->CO2   + H         }      &      1.700(-33)  &  0    &  1.510(3) & \ce{CO    + O    + H2  ->CO2   + H2        }      &      1.700(-33)  &  0    &  1.510(3)  \\
\ce{H2    + H          ->H     + H    + H  }      &      1.000(-08)  &  0    &  8.410(4) & \ce{H2    + H2         ->H     + H    + H2 }      &      1.000(-08)  &  0    &  8.410(4)  \\
\ce{CH    + H          ->C     + H    + H  }      &      6.000(-09)  &  0    &  4.020(4) & \ce{CH    + H2         ->C     + H    + H2 }      &      6.000(-09)  &  0    &  4.020(4)  \\
\ce{CH2   + H          ->C     + H2   + H  }      &      5.000(-10)  &  0    &  3.260(4) & \ce{CH2   + H2         ->C     + H2   + H2 }      &      5.000(-10)  &  0    &  3.260(4)  \\
\ce{CH2   + H          ->CH    + H    + H  }      &      1.560(-08)  &  0    &  4.488(4) & \ce{CH2   + H2         ->CH    + H    + H2 }      &      1.560(-08)  &  0    &  4.488(4)  \\
\ce{CH3   + H          ->CH2   + H    + H  }      &      1.700(-08)  &  0    &  4.560(4) & \ce{CH3   + H2         ->CH2   + H    + H2 }      &      1.700(-08)  &  0    &  4.560(4)  \\
\ce{CH3   + H          ->CH    + H2   + H  }      &      1.100(-08)  &  0    &  4.280(4) & \ce{CH3   + H2         ->CH    + H2   + H2 }      &      1.100(-08)  &  0    &  4.280(4)  \\
\ce{CH4   + H          ->CH3   + H    + H  }      &      7.500(-07)  &  0    &  4.570(4) & \ce{CH4   + H2         ->CH3   + H    + H2 }      &      7.500(-07)  &  0    &  4.570(4)  \\
\ce{NH3   + H          ->NH    + H2   + H  }      &      3.100(-08)  &  0    &  4.686(4) & \ce{NH3   + H2         ->NH    + H2   + H2 }      &      3.100(-08)  &  0    &  4.686(4)  \\
\ce{NH3   + H          ->NH2   + H    + H  }      &      4.170(-08)  &  0    &  4.720(4) & \ce{NH3   + H2         ->NH2   + H    + H2 }      &      4.170(-08)  &  0    &  4.720(4)  \\
\ce{OH    + H          ->O     + H    + H  }      &      6.000(-09)  &  0    &  5.090(4) & \ce{OH    + H2         ->O     + H    + H2 }      &      6.000(-09)  &  0    &  5.090(4)  \\
\ce{H2O   + H          ->OH    + H    + H  }      &      5.800(-09)  &  0    &  5.290(4) & \ce{H2O   + H2         ->OH    + H    + H2 }      &      5.800(-09)  &  0    &  5.290(4)  \\
\ce{O2    + H          ->O     + O    + H  }      &      6.000(-09)  &  0    &  5.230(4) & \ce{O2    + H2         ->O     + O    + H2 }      &      6.000(-09)  &  0    &  5.230(4)  \\
\ce{CO    + H          ->C     + O    + H  }      &      1.480(-04)  & -3.1  &  1.290(5) & \ce{CO    + H2         ->C     + O    + H2 }      &      1.480(-04)  & -3.1  &  1.290(5)  \\
\ce{HCO   + H          ->CO    + H    + H  }      &      6.600(-11)  &  0    &  7.820(3) & \ce{HCO   + H2         ->CO    + H    + H2 }      &      6.600(-11)  &  0    &  7.820(3)  \\
\ce{H2CO  + H          ->HCO   + H    + H  }      &      2.450(-08)  &  0    &  3.805(4) & \ce{H2CO  + H2         ->HCO   + H    + H2 }      &      2.450(-08)  &  0    &  3.805(4)  \\
\ce{H2CO  + H          ->CO    + H2   + H  }      &      1.420(-08)  &  0    &  3.210(4) & \ce{H2CO  + H2         ->CO    + H2   + H2 }      &      1.420(-08)  &  0    &  3.210(4)  \\
\ce{CN    + H          ->C     + N    + H  }      &      4.200(-10)  &  0    &  7.100(4) & \ce{CN    + H2         ->C     + N    + H2 }      &      4.200(-10)  &  0    &  7.100(4)  \\
\ce{HCN   + H          ->CN    + H    + H  }      &      2.150(-04)  & -2.6  &  6.280(4) & \ce{HCN   + H2         ->CN    + H    + H2 }      &      2.150(-04)  & -2.6  &  6.280(4)  \\
\ce{NO    + H          ->N     + O    + H  }      &      1.600(-09)  &  0    &  7.460(4) & \ce{NO    + H2         ->N     + O    + H2 }      &      1.600(-09)  &  0    &  7.460(4)  \\
\ce{SO2   + H          ->SO    + O    + H  }      &      4.200(-10)  &  0    &  5.540(4) & \ce{SO2   + H2         ->SO    + O    + H2 }      &      4.200(-10)  &  0    &  5.540(4)  \\
\hline
    \end{tabular}
    \tablecomments{$x(y) \equiv x \times 10^y$; a is in units of cm$^3$ s$^{-1}$ or cm$^6$ s$^{-1}$ for two and three body reaction respectively. }
    \label{tab:react}
\end{table}

\section{Disk structure variations}
\label{app:tgasthick}

\begin{figure*}

    \begin{minipage}{0.499\hsize}
    \includegraphics[width = \hsize]{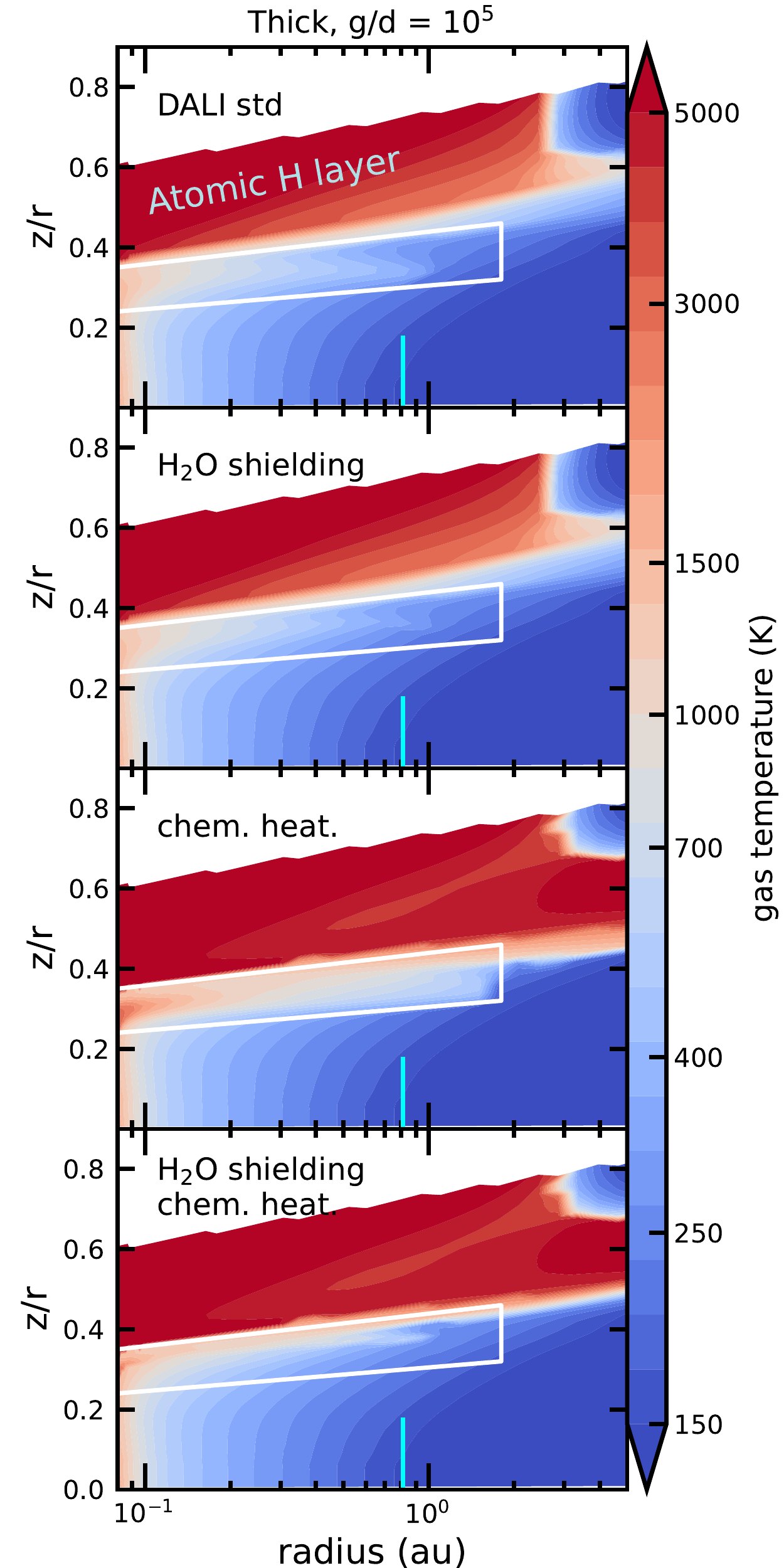}
    \end{minipage}%
    \begin{minipage}{0.499\hsize}
    \includegraphics[width = \hsize]{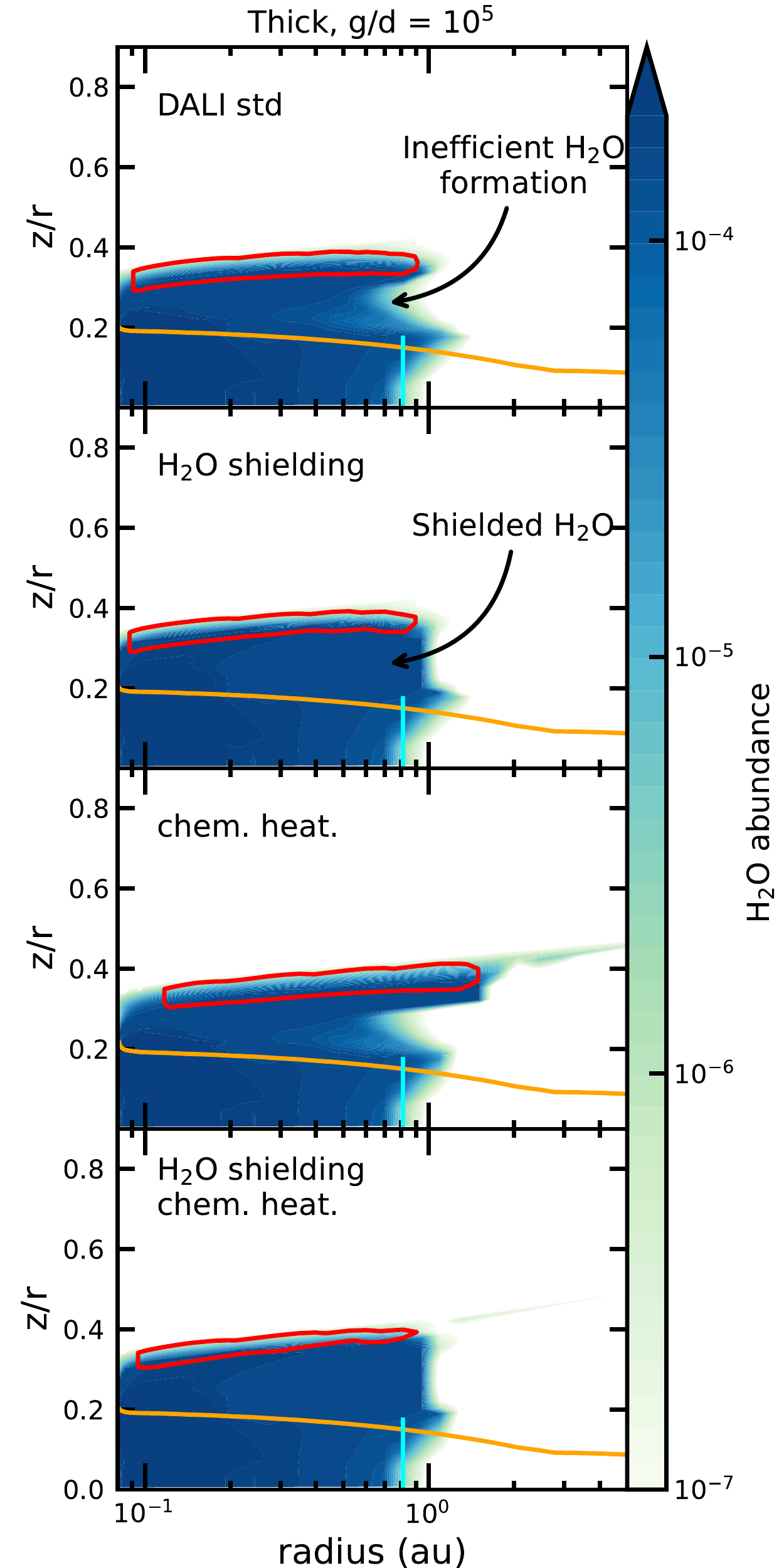}
    \end{minipage}
    \caption{Same as Fig.~\ref{fig:Tgas_chem} but for the thick model with a surface layer gas-to-dust ratio of $10^5$.}
    \label{fig:Tgas_chem_thick}
    \label{fig:Water_abu_thickE5}
\end{figure*}

Figure~\ref{fig:Tgas_chem_thick} shows the gas temperature and water abundance for a thick model with a surface layer gas-to-dust ratio of $10^5$. The more puffed up structure intercepts more stellar flux leading to a warmer disk in general. Pushing out the mid-plane water ice-line as well as the water emitting region. Furthermore, the UV-heated molecular and water emitting layers are higher up in the disk.

Apart from the upward and outward scaling there are no significant differences in the water abundance structure, the gap in the water abundance at large radii also appears in the thicker model. As in the thin mode the inclusion of water UV-shielding strongly increases the abundance of this region, making water the dominant water carrier in the entire inner disk region.

Increasing the amount of small dust has the main effect of cooling down the surface layers. Especially the UV heated layer shrinks radially and move upward. This is mostly caused by the more efficient gas-grain cooling. This directly effects the water abundance structure. The water is less extended in the surface layers due to the lower temperatures. This naturally causes a smaller emitting area for the water.
\section{Spectra for disk variations}
\label{app:dust}

\begin{figure*}
    \centering
    \begin{minipage}{0.49\hsize}
    \includegraphics[width = \hsize]{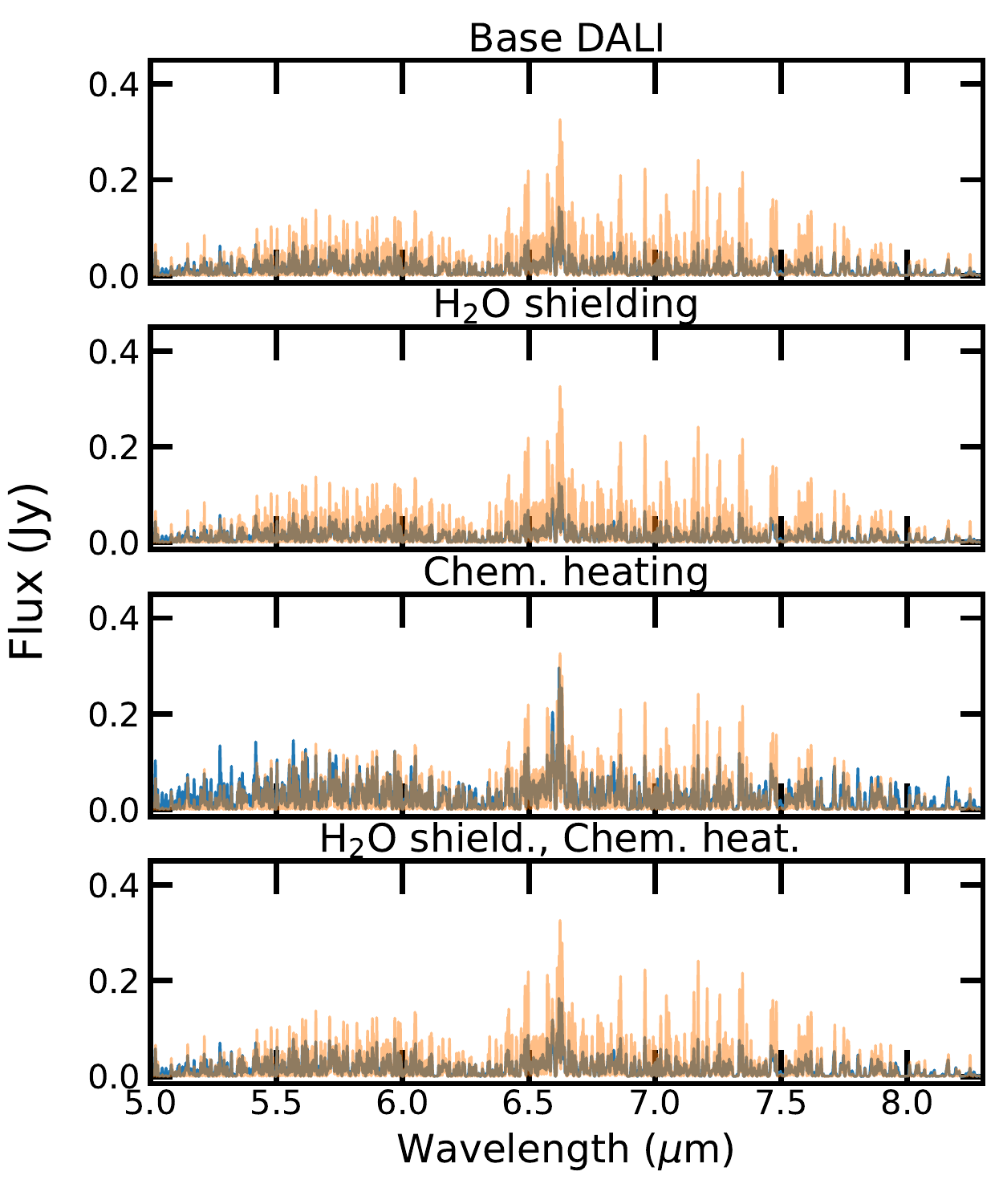}
    \end{minipage}%
    \begin{minipage}{0.49\hsize}
    \includegraphics[width = \hsize]{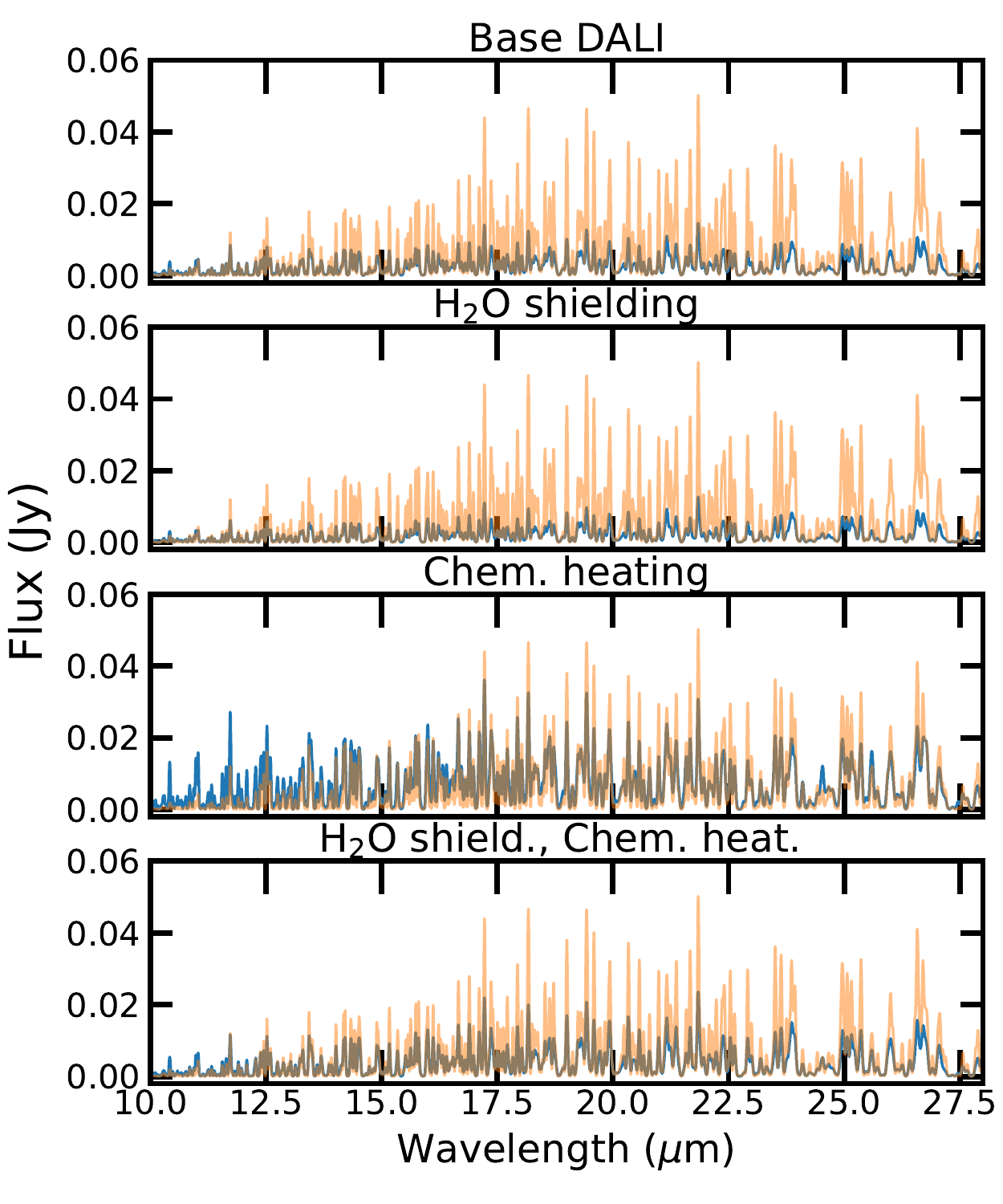}
    \end{minipage}

    \caption{Same as Fig.~\ref{fig:Spectra_thinE5} but for the thin model with a surface layer gas-to-dust ratio of $10^4$. }
    \label{fig:Spectra_thinE4}
\end{figure*}

\begin{figure*}
    \centering
    \begin{minipage}{0.49\hsize}
    \includegraphics[width = \hsize]{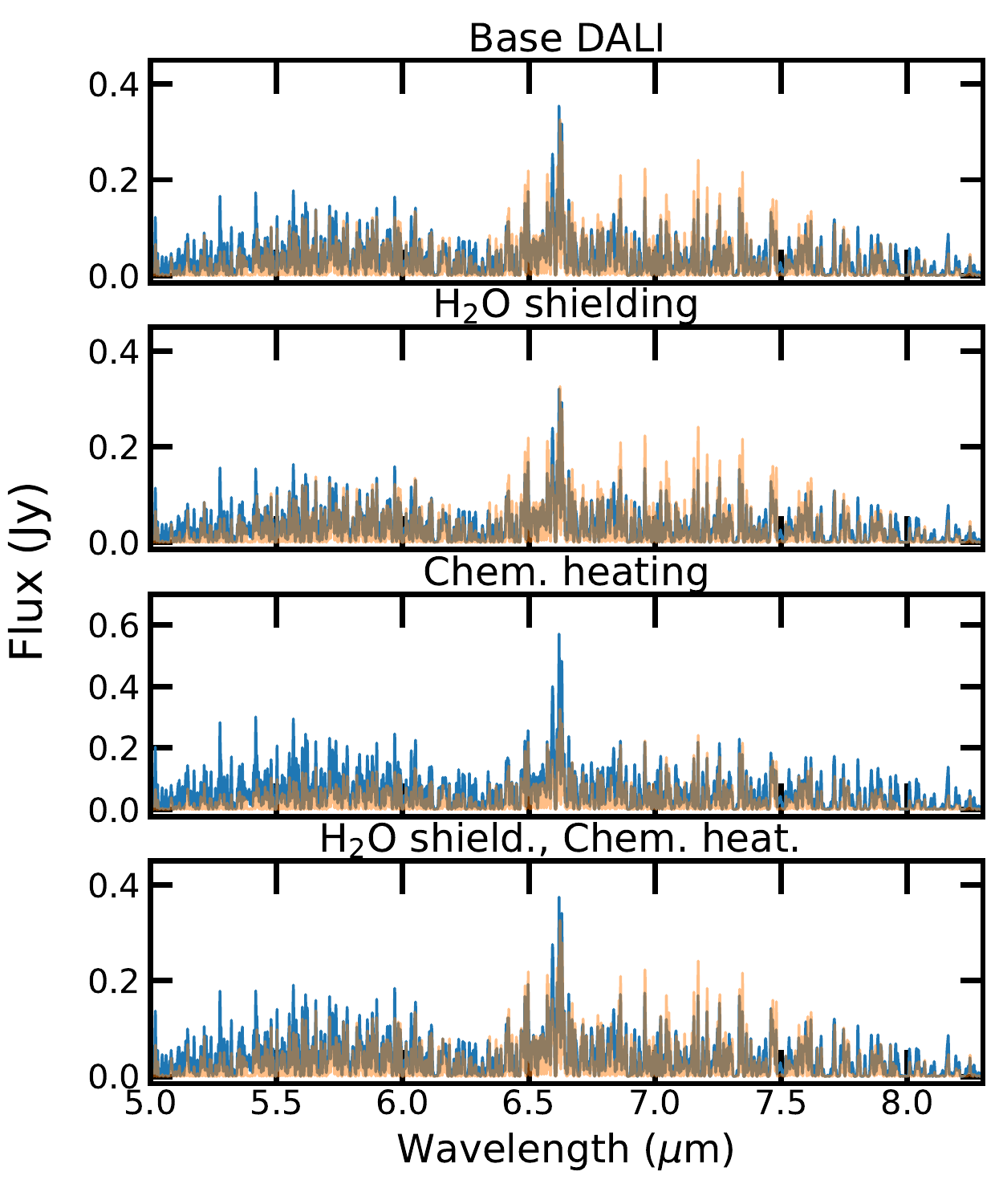}
    \end{minipage}%
    \begin{minipage}{0.49\hsize}
    \includegraphics[width = \hsize]{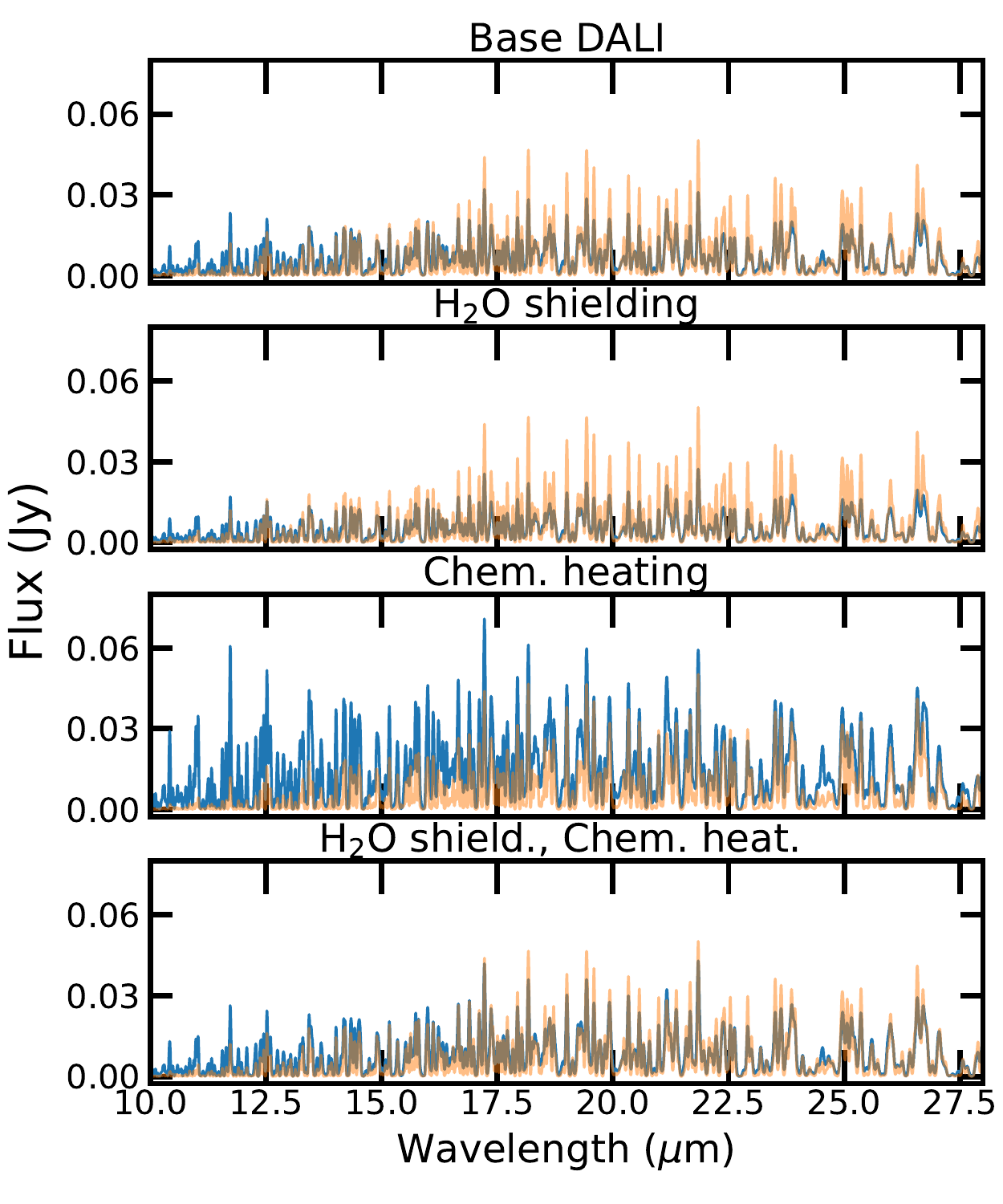}
    \end{minipage}

    \caption{Same as Fig.~\ref{fig:Spectra_thinE5} but for the thick model with a surface layer gas-to-dust ratio of $10^4$. }
    \label{fig:Spectra_thickE4}
\end{figure*}

\begin{figure*}
    \centering
    \begin{minipage}{0.49\hsize}
    \includegraphics[width = \hsize]{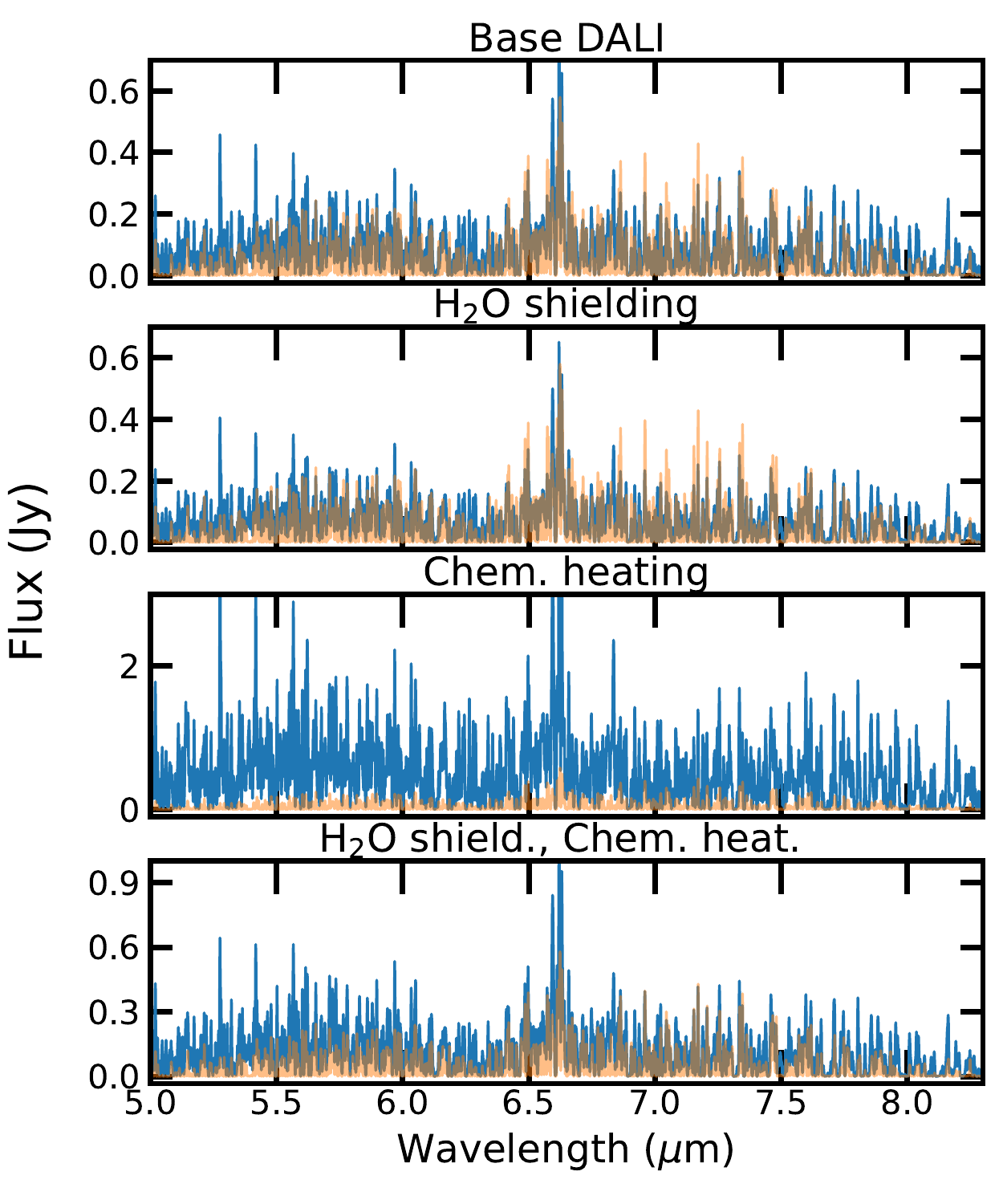}
    \end{minipage}%
    \begin{minipage}{0.49\hsize}
    \includegraphics[width = \hsize]{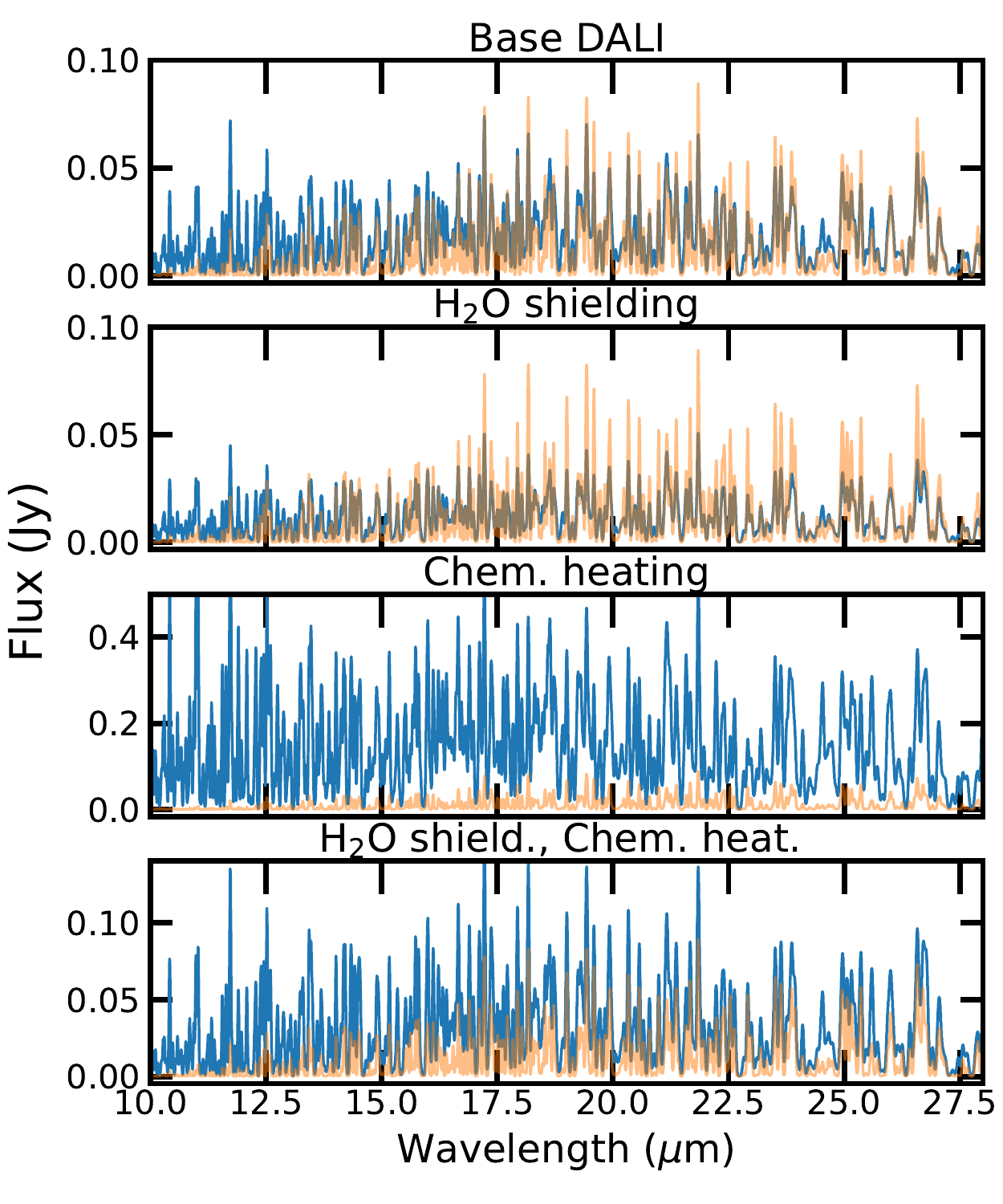}
    \end{minipage}

    \caption{Same as Fig.~\ref{fig:Spectra_thinE5} but for the thick model with a surface layer gas-to-dust ratio of $10^5$. }
    \label{fig:Spectra_thickE5}
\end{figure*}

Figures~\ref{fig:Spectra_thinE4},~\ref{fig:Spectra_thickE4}~and~\ref{fig:Spectra_thickE5} show the water spectra for three structures in Fig.~\ref{fig:dust_dens_temp}, the final spectrum is in Fig.~\ref{fig:Spectra_thinE5} in the main text. The behavior of the spectra with extra chemical heating as well as with water UV-shielding is consistent over the different structures. This implies that our observations are robust to disk structure. 

Water line flux is increased with increased disk scaleheight as well as with increased gas-to-dust ratios. Lower amounts of dust allow for more UV photons to penetrate the \ce{H2O} emitting layer, increasing the gas-temperature. In the more puffed-up models, the \ce{H2O} emitting layer is present at slightly higher temperatures as lower densities suppress gas cooling. 

For the models that include water UV-shielding, it is clear that the thin model with a $10^4$ surface gas-to-dust ratio undershoots the slab model, and that the thick model with a $10^4$ surface gas-to-dust ratio only just matches the flux when extra chemical heating is included. As such a surface layer gas-to-dust ratio $>10^4$ or very thick disk (h/r$>$0.1 at 1 au) is necessary to reproduce the water spectra. 

The model with excess chemical heating, absorbs most of the incident UV photon energy and deposits it in the gas. Increasing the gas-heating beyond this needs a different energy source than the central star. One obvious candidate would be heating from accreting gas \citep[e.g.][]{Glassgold2004}. As the temperature is driven by the local heating due to stellar photon and gas cooling, local heating is going to be far more effective in increasing the temperature than heating near the disk mid-plane. Inclusion of accretion heating would allow for lower gas-to-dust ratios in the disk surface to still reproduce the observed spectra. A scenario with a lower gas-to-dust ratio, but with accretion heating to compensate would be hard to distinguish from a high gas-to-dust ratio with no accretion heating in the water emission. The lack of a deeper, colder visible gas reservoir could be seen in various less abundant species and help distinguish amongst these scenarios.

\end{document}